\documentclass[final,1p,times]{elsarticle} 
\usepackage{graphicx} 
\usepackage{amssymb} 
\usepackage{amsthm} 
\usepackage{lineno} 

\journal{Nuclear Physics A} 
\begin{document} 

\begin{frontmatter} 


\title{Quarkonia measurements with ALICE at the LHC}

\author{Magdalena Malek$^{a}$ for the ALICE collaboration}

\address[a]{Institut de Physique Nucl\'{e}aire d'Orsay (IPNO) - France\\CNRS: UMR8608 - IN2P3 - Universit\'{e} Paris Sud - Paris XI}
\begin{abstract} 
In this paper, we summarize the perspectives on quarkonia detection in the ALICE experiment at the
CERN LHC, both in the dielectron and dimuon decay channels.

\end{abstract} 

\end{frontmatter} 



\section{Introduction}\label{Intro}
Quarkonia production in heavy ion collisions is considered to be one of the most powerful tools to
probe the quark-gluon plasma (QGP) formation. In fact, the in-medium behavior of the J/$\psi$ was
proposed to test whether deconfinement has occurred. It was predicted that in a deconfined medium,
color screening dissolves the $c\overline{c}$ bound state~\cite{JPsiSatz}. This result provided a
strong motivation for experimental studies of quarkonia production at SPS and RHIC.\\
\indent ALICE (A Large Ion Collider Experiment)~\cite{PPR1, PPR2, ALICE} is the experiment
dedicated to the study of heavy ion collisions in the unprecedented energy regime of the LHC
(Large Hadron Collider) collider. Its goal is to study the properties of deconfined matter: the
Quark Gluon Plasma (QGP). The LHC will collide Pb-Pb at $\sqrt{s_{NN}}$ = 5.5 TeV opening up new
perspectives in the QGP study. To extract information about the QGP from Pb-Pb collisions, the
comparison to less complex systems, for which deconfinement effects are not expected, is
mandatory. The LHC will deliver p-p collisions ($\sqrt{s_{NN}}$=14 TeV) and p-Pb collisions
($\sqrt{s_{NN}}$ = 8.8 TeV)
providing a solid baseline for the Pb-Pb system.\\
\indent The ALICE detector is composed of a central barrel system ($\mid\eta\mid<$0.9), a muon
spectrometer (-4.0$<\eta<$-2.5) and several small additional detectors. Quarkonia will be measured
in ALICE via the dielectron channel at midrapidity and via the dimuon channel at forward rapidity.
In the following, we report on a selection of results from analyses which are being prepared on
simulations.




\section{Quarkonia detection in the dielectron channel}\label{detEl}
The measurement of electron pairs in the ALICE central barrel is provided by the combination of
several detectors that are described here, as seen by a particle travelling out from the
interaction point:
\begin{itemize}
\item Inner Tracking System (ITS)~\cite{TDR_ITS} allows 3-D reconstruction of the primary vertex,
secondary vertex finding, and particle identification via $dE/dx$. It is composed of three
subsystems of two layers each: the Silicon Pixel Detector, the Silicon Drift Detector and the
Silicon Strip Detector.
\item Time Projection Chamber (TPC)~\cite{TDR_TPC}, optimized for large multiplicity environments,
 allows track finding, momentum measurement, and charged hadron identification via $dE/dx$. The
detector has an inner radius of 0.9 m, an outer radius of 2.5 m and a length of 5.1 m. The
momentum resolution for the track reconstruction, including TPC and ITS information, is expected
to be better than 2\% for p$_{t}$~$<$~20~GeV/c.
\item Transition Radiation Detector (TRD)~\cite{TDR_TRD} allows electron identification and also provides
fast triggering. Electron identification is provided by the TRD for momenta larger than 1~GeV/c.
This detector is made of 18 longitudinal supermodules, 6 radial layers, and 5 stacks along the
beam axis.
\end{itemize}

The expected early statistic collected by ALICE for quarkonia measurements is summarized in
Table~1. These numbers correspond to 10$^{6}$ s data taking period of central (10\% most central)
Pb-Pb collisions at a luminosity of 5~$\times$~10$^{26}$ cm$^{-2}$s$^{-1}$ and for a charged
particle multiplicity density dN$_{ch}$/d$\eta$~=~3000. The invariant mass resolution for the
quarkonia was studied in the full simulation. The reconstructed peaks were fitted by a Gaussian
and the invariant mass resolution for the J/$\psi$ ($\Upsilon$) is found to be
30~(90)~MeV/c$^{2}$. The analysis showed that the J/$\psi$ signal can be reconstructed up to
p$_{t}$~=~10~GeV/c.

\begin{table}[h]
\begin{center}
 \begin{tabular}{c|c|c|c|c}
    \hline
    state  &S(x10$^{3}$)&B(x10$^{3}$)&S/B&S/$\sqrt{S+B}$\\
    \hline
    \hline
    J/$\psi$&121.1&88.2&1.4&265\\
    $\Upsilon$&1.3&0.8&1.6&28\\
    $\Upsilon^{'}$&0.46&0.8&0.6&13\\
    \hline

    \end{tabular}
    \caption{Expected signal rates (S), background (B), signal-to-background ratios and
    significance, in the dielectron channel, for charmonia and bottomonia states integrated over full
    acceptance with an interval of $\pm$1.5$\sigma$
    around each resonance mass for Pb-Pb collisions.
    All yields correspond to annual data taking (10$^{6}$ s)
    at a luminosity of 5~$\times$~10$^{26}$ cm$^{-2}$s$^{-1}$.}
    \end{center}
    \label{tab1}
    \end{table}
\section{Quarkonia detection in the dimuon channel}\label{detMu}
The detection of muon pairs in the forward rapidity region is provided by the muon
spectrometer~\cite{TDR_MUON}. This detector contains a set of absorbers: a front absorber, a muon
filter, a beam shielding and an absorber against the LHC background. The goal of these absorbers
is to suppress hadron and electron background. The tracking system that allows the muon trajectory
reconstruction is composed of five tracking stations with two detector planes each. The dipole
magnet with a field integral of 3 Tm along the beam axis provides the bending power to measure the
momenta of muons. Two trigger stations provide a fast electronic signal for the trigger selection
of muon events. The p$_{t}$ cut of 1~(2)~GeV/c applied to single muons allows charmonia
(bottomonia) detection down to zero transverse momenta. The invariant mass resolution for the
J/$\psi$ ($\Upsilon$) is expected to be around 70~(100)~MeV/c$^{2}$. The high-p$_{t}$ reach for
the J/$\psi$ is expected to be around 20~GeV/c. The expected quarkonia signal and background
yields, and the corresponding signal-to-background ratios and significance for the 5\% most
central collisions are presented in Table~2.
\begin{table}[h]
\begin{center}
\begin{tabular}{c|c|c|c|c}
\hline
    state  &S(x10$^{3}$)&B(x10$^{3}$)&S/B&S/$\sqrt{S+B}$\\
    \hline
    \hline
    J/$\psi$&130&680&0.20&150\\
$\psi^{'}$&3.7&300&0.01&6.7\\
$\Upsilon$&1.3&0.8&1.7&29\\
$\Upsilon^{'}$&0.35&0.54&0.65&12\\
$\Upsilon^{''}$&0.20&0.42&0.48&8.1\\
\hline
\end{tabular}
\caption{Expected signal rates (S), background (B), signal-to-background ratios and
    significance, in the dimuon channel, for charmonia and bottomonia states for annual running time of Pb-Pb collisions at a luminosity
    of 5$\times$10$^{26}$~cm$^{-2}$s$^{-1}$. The yields correspond to an interval of
    $\pm$2$\sigma$ around the resonance mass.}
\end{center}
    \label{tab2}
    \end{table}

\section{Prompt and secondary J/$\psi$} In addition to prompt J/$\psi$ and $\psi^{'}$ also those
from B decays have to be taken into account. The contribution of J/$\psi$ ($\psi^{'}$) from B
decay is about 22~(39)~\% of the total J/$\psi$ ($\psi^{'}$) yields. To separate the prompt
J/$\psi$'s from the secondary ones, dielectron pairs with a displaced vertex must be identified.
In fact, the J/$\psi$ originating from B decay are produced at large distances (several hundred of
microns) from the primary vertex. The vertexing capabilities of the ITS allow to apply this method
in the central barrel. This analysis is not expected to be performed in the forward rapidity
region because the muon spectrometer does not provide vertexing capabilities at present.

\section{Suppression studies} Two extreme suppression scenarios were studied. The first one
characterized by a high critical deconfinement temperature at T$_{c}$~=~270~MeV~\cite{SUPP_h} and
the second one using T$_{c}$~=~190~MeV~\cite{SUPP_l}. The ratios of the resonance rates over those
for beauty as a function of the number of participants were studied to check the detector ability
to distinguish between different suppression scenarios. It was shown~\cite{PPR2} that the error
bars for J/$\psi$ and $\Upsilon$ to open beauty ratio are small enough to distinguish between
these two suppression scenarios.


\section{Quarkonia polarization} Quarkonia polarization measurements should allow to distinguish
different production mechanisms, since different models predict different polarizations. Quarkonia
polarization can be reconstructed from the angular distribution of the decay products (dimuons or
dielectrons) in the quarkonia rest frame. It has been predicted that an increase of J/$\psi$
polarization may be expected if the QGP is formed~\cite{Polar}.

\begin{figure}[ht]
\centering
\includegraphics[scale=0.4]{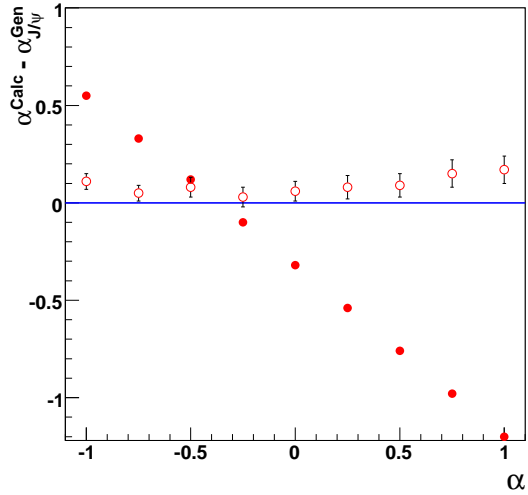}
\caption[]{The difference between the reconstructed polarization ($\alpha_{Calc}$) and generated
one ($\alpha_{Gen}$) as a function of $\alpha$ for the J/$\psi$ in Pb-Pb collisions at
$\sqrt{s_{NN}}$~=~5.5 TeV. The full circles correspond to the case without background subtraction
(S/B=0.2). The open circles represent the case where the background was subtracted.}

\label{polar}
\end{figure}

It is expected that a measurement of the J/$\psi$ and $\Upsilon$ polarization (integrated over
centrality) will be possible after an annual data taking of Pb-Pb collision. It can be seen in
Fig.~\ref{polar} that the background subtraction is mandatory to obtain a correct estimation of
the J/$\psi$ polarization.



\section{Summary}
We presented an overview of the ALICE perspectives for quarkonia physics. We conclude that the
quarkonia are interesting tools to probe the properties of the strongly-interacting medium formed
in heavy ion collisions. The ALICE detector can measure quarkonia in the dielectron and dimuon
channels in different rapidity domains. It can detect quarkonia down to p$_{t}$~=~0 in both
channels. The excellent tracking, vertexing and particle identification capabilities of ALICE will
allow us to explore this very rich topic. It is believed that the large statistics of quarkonia
expected at the LHC and the experimental capabilities of the ALICE detector will help to clarify
the quarkonia production picture that is still today one of the most discussed topics in heavy-ion
physics.


\end{document}